\documentclass[a4paper,11pt]{article}
\usepackage[utf8]{inputenc}
\usepackage{default}
\usepackage{color}
\usepackage{amssymb}
\usepackage{amsmath}
\usepackage{graphicx}
\usepackage{mathtools}
\usepackage{ragged2e}
\usepackage{hyperref}
\hypersetup{
    colorlinks=true,
    linkcolor=blue,
    filecolor=cyan, 
    urlcolor=red,
    citecolor=red,
} 
\usepackage[titletoc,toc,title]{appendix}
\numberwithin{equation}{section}
\usepackage{cleveref}
\usepackage[margin=2.5cm]{geometry}
\usepackage{cite}
\usepackage{epsfig}
\usepackage{float}
\usepackage{cancel}
\usepackage{amsfonts}
\usepackage{enumitem}
\usepackage[font={footnotesize,it}]{caption}
\usepackage{authblk}

\begin{document}
\begin{titlepage}
\title{Holographic Entanglement Negativity for Conformal Field Theories with a Conserved Charge}
\date{}

\author[1,2,3]{Parul Jain\thanks{\noindent E-mail:~ parul.jain@ca.infn.it}}
\affil[1]{Dipartimento di Fisica, Universit\`a di Cagliari\\Cittadella Universitaria, 09042 Monserrato, Italy\smallskip}
\affil[2]{INFN, Sezione di Cagliari, Italy\bigskip}
\author[3]{Vinay Malvimat\thanks{\noindent E-mail:~ vinaymm@iitk.ac.in}} 
\author[3]{Sayid Mondal\thanks{\noindent E-mail:~  sayidphy@iitk.ac.in}}
\author[3]{Gautam Sengupta\thanks{\noindent E-mail:~  sengupta@iitk.ac.in}}

\affil[3]{
Department of Physics\\

Indian Institute of Technology\\ 

Kanpur, 208016\\ 

India}

\maketitle
\begin{abstract}
\noindent
\justify

We study the application of our recent holographic entanglement negativity conjecture
for mixed states of adjacent subsystems in conformal field theories with a conserved charge.
In this context we obtain the holographic entanglement negativity for zero and finite temperature mixed state configurations in $d$-dimensional conformal field theories dual to bulk extremal and non extremal charged $AdS_{d+1}$ black holes. Our results conform to quantum information theory expectations and constitute significant consistency checks for our conjecture.

\end{abstract}
\end{titlepage}
\tableofcontents
\pagebreak

\section{Introduction}
\label{sec1}
\justify

In recent times quantum entanglement has emerged as an important facet of modern fundamental physics, relating diverse fields ranging from many body theory to issues of quantum gravity and black holes. In this context the measure of {\it entanglement entropy} has played a crucial role in the characterization of quantum entanglement for bipartite pure states. In quantum information theory the entanglement entropy is defined as the von Neumann entropy of the reduced density matrix for the corresponding subsystem. Significantly, the entanglement entropy may be computed through a replica technique for bipartite states in  $(1+1)$-dimensional conformal field theories ($CFT_{1+1}$)  as described in \cite{Calabrese:2004eu,Calabrese:2009qy}. Interestingly
Ryu and Takayanagi  in a seminal work \cite{Ryu:2006bv,Ryu:2006ef,Nishioka:2009un} proposed an elegant holographic entanglement entropy conjecture for bipartite states of dual $d$-dimensional conformal field theories ($CFT_d$) in the framework of the $AdS/CFT$ correspondence. The Ryu-Takayanagi conjecture inspired extensive investigations in various aspects of entanglement in holographic $CFT$s \cite{Takayanagi:2012kg,Nishioka:2009un,e12112244,Blanco:2013joa,Fischler2013,Fischler:2012uv,Chaturvedi:2016kbk} (and references therein). A proof of this conjecture from a bulk perspective was subsequently developed in a series of communications, first for the $AdS_3/CFT_2$ scenario and later extended to a generic $AdS_{d+1}/CFT_{d}$ framework \cite{Fursaev:2006ih,Headrick:2010zt,Hartman:2013mia, Faulkner:2013yia,Casini:2011kv,Lewkowycz:2013nqa}.

It is well known however in quantum information theory that entanglement entropy fails to characterize mixed state entanglement as it receives contributions which are irrelevant to the entanglement of the configuration in question. 
Hence characterization of mixed state entanglement was a complex and subtle issue which required the introduction of suitable measures. In a seminal work Vidal and Werner \cite{PhysRevA.65.032314} addressed this critical issue and proposed a computable measure for characterizing the upper bound on the distillable entanglement for bipartite mixed states, termed as {\it entanglement negativity}. It could be shown that this measure is non convex and an entanglement monotone \cite{Plenio:2005cwa}. Interestingly, in a series of communication the authors in \cite{Calabrese:2012nk,Calabrese:2012ew,Calabrese:2014yza} computed the entanglement negativity for several bipartite mixed state configurations in 
$CFT_{1+1}$s employing a suitable replica technique.

The above discussion naturally leads to the issue of a holographic characterization for the entanglement negativity of bipartite pure and mixed states in dual $CFT$s, in terms of the bulk geometry through the $AdS/CFT$ correspondence. There were several attempts
in the literature \cite{Rangamani:2014ywa,Perlmutter:2015vma} to address this issue and despite significant progress a clear elucidation of a holographic characterization for the entanglement negativity remained a crucial open problem. In the recent past, two of the present authors (VM and GS) in the collaboration \cite{Chaturvedi:2016rcn,Chaturvedi:2016rft,Chaturvedi:2016opa} (CMS), proposed a holographic entanglement negativity conjecture for bipartite states in the dual $CFT$s.
According to their conjecture, the holographic negativity characterizing the entanglement of a simply connected single subsystem with the rest of the system, is described by a specific algebraic sum of the areas of co dimension two bulk static minimal surfaces (lengths of space like geodesics in the $AdS_3/CFT_2$ scenario) anchored on appropriate subsystems. In the $AdS_3/CFT_2$ context \cite{Chaturvedi:2016rcn} their conjecture could exactly reproduce the universal part of the corresponding $CFT_{1+1}$ replica technique results, in the large central charge limit. Furthermore their analysis was strongly confirmed through a large central charge analysis of the entanglement negativity in $CFT_{1+1}$ employing the monodromy technique in \cite{Malvimat:2017yaj}. The corresponding higher dimensional extension of the conjecture was substantiated through strong consistency checks involving applications to specific examples \cite{Chaturvedi:2016rft}. Interestingly, this reproduced certain universal features of the holographic entanglement negativity for the corresponding $AdS_3/CFT_2$ scenario \cite{Chaturvedi:2016rcn}. However we should mention here that a formal bulk proof for their conjecture along the lines of \cite{Lewkowycz:2013nqa} remains a non trivial open issue which needs to be addressed.

Recently, in a subsequent communication \cite{Jain:2017aqk} the present authors proposed an independent holographic entanglement negativity conjecture for a bipartite mixed state configuration of adjacent intervals in a dual $CFT_{1+1}$.
The conjecture involved a specific algebraic sum of the lengths of space like geodesics in the dual bulk $AdS_3$ configuration which are anchored on appropriate intervals. Interestingly, this reduced to the holographic mutual information between the two intervals upto a numerical constant \footnote{\label{fn1} Note that  this matching between the universal part of the entanglement negativity and the mutual information for the case of adjacent intervals has also been reported for time dependent situations following both local and global quenches in a $CFT_{1+1}$  \cite{Coser:2014gsa,Wen:2015qwa}}. Remarkably, as earlier \cite{Chaturvedi:2016rcn,Chaturvedi:2016rft,Chaturvedi:2016opa}, 
in this case also the holographic entanglement negativity exactly reproduced the universal part of the corresponding $CFT_{1+1}$ results in the large central charge limit. 

A higher dimensional extension of the above conjecture for the mixed state of adjacent subsystems in a holographic $CFT_d$ was proposed subsequently in \cite{Jain:2017xsu}. As earlier this involved a specific algebraic sum of the areas of co-dimension two bulk static minimal surfaces anchored on the respective subsystems in the dual $CFT_d$. This extension was substantiated through applications to specific higher dimensional examples constituting strong consistency checks for the holographic conjecture. These involved the computation of the holographic entanglement negativity for mixed states of adjacent subsystems described by rectangular strip geometries in $CFT_d$s dual to bulk pure $AdS_{d+1}$ geometry and  $AdS_{d+1}$-Schwarzschild black hole. Quite interestingly, for the finite temperature case involving the dual $AdS_{d+1}$-Schwarzschild black hole, the holographic entanglement negativity scales as the area of the entangling surface in a high temperature approximation. Note that this is unlike the case of entanglement entropy which scales as the volume of the subsystem at high temperatures \cite{Fischler:2012ca}. For the holographic entanglement negativity on the other hand, all volume dependent thermal terms cancel out leading to a purely area dependent expression . This conforms to the standard quantum information theory expectations for the entanglement negativity measure. Interestingly, the area law for entanglement negativity  has also been reported for condensed matter system such as the  finite temperature quantum spin model and the two dimensional harmonic lattice \cite{DeNobili:2016nmj,PhysRevE.93.022128}. Subsequently, a covariant version of the holographic entanglement negativity conjecture described in \cite{Jain:2017aqk}, was proposed in \cite{Jain:2017uhe} for time dependent mixed state configurations of adjacent intervals in a dual $CFT_{1+1}$. 

In this article we further substantiate the higher dimensional $AdS_{d+1}/CFT_d$ extension of the holographic entanglement negativity conjecture described above, with additional non trivial consistency checks involving the application to distinct 
examples. This involves zero and finite temperature mixed state configurations of adjacent subsystems with rectangular strip geometries in holographic $CFT_d$s with a conserved charge, dual to bulk extremal and non extremal RN-$AdS_{d+1}$ black holes. Unlike for the case of $CFT_d$s dual to $AdS_{d+1}$-Schwarzschild black holes \cite{Jain:2017xsu}, the holographic entanglement negativity for the present case necessitates 
perturbative expansions involving non trivial limits of the relevant parameters (see also  \cite{Kundu:2016dyk,Blanco:2013joa} for the corresponding case of entanglement entropy). In order to illustrate this we initially consider the $AdS_4/CFT_3$ examples for simplicity and subsequently describe the more general $AdS_{d+1}/CFT_d$ scenario. 

In this context we first compute the holographic entanglement negativity for bipartite mixed states of adjacent subsystems in $CFT_3$s dual to bulk non-extremal and extremal RN-$AdS_4$ black holes. We demonstrate that the holographic entanglement negativity following from our conjecture in the various limits of the relevant parameters conform to quantum information expectations.  Hence these serve as significant consistency checks for the universality of our conjecture although a bulk proof remains an outstanding open issue. The corresponding $AdS_{d+1}/CFT_d$ case necessitates the perturbative description of the holographic entanglement negativity involving various limits of a distinct set of parameters. However the results of this exercise are similar to the previous case of $AdS_4/CFT_3$ and lead to identical conclusions in the appropriate limits for the relevant parameters.

The article is organized as follows. In section \ref{sec2} we describe our holographic entanglement negativity conjecture for mixed state configurations of adjacent subsystems characterized by a rectangular strip geometry in $CFT_d$s dual to bulk $AdS_{d+1}$  configurations. Subsequently, in section \ref{sec3} we compute the holographic entanglement negativity for mixed states of adjacent subsystems in the $AdS_4/CFT_3$ scenario. In section \ref{sec4} we obtain the holographic entanglement negativity for the required mixed states in
the $AdS_{d+1}/CFT_d$ scenario. In the final section \ref{sec5} we present a summary of our results and conclusions.

\section{Holographic entanglement negativity conjecture}\label{sec2}
In this section we briefly review the holographic entanglement negativity conjecture  for a bipartite mixed state configuration of adjacent subsystems in dual $CFT$s. To this end we first describe the holographic entanglement negativity conjecture for the mixed state configuration above in the context of the $AdS_3/CFT_2$ scenario \cite{Jain:2017aqk}. Following this we briefly discuss the extension of our conjecture to a generic higher dimensional $AdS_{d+1}/CFT_d$ scenario \cite{Jain:2017xsu}. 

The entanglement negativity for a bipartite mixed state configuration in a $CFT_{1+1}$ may be obtained through a suitable replica technique as described in \cite {Calabrese:2012ew,Calabrese:2012nk,Calabrese:2014yza}. This involves the spatial tripartition of the $CFT_{1+1}$ into the intervals $A_1$ and  $A_2$ such that $ A=A_1\cup A_2$, and the rest of the system is  $A^c=(A_1\cup A_2)^c$. The entanglement negativity is then defined as
\begin{equation}\label{cftneg}
\mathcal{E} = \lim_{n_e \rightarrow 1 } \ln \mathrm{Tr}(\rho_A^{T_2})^{n_e},
\end{equation}
where $\rho_A^{T_2}$ is the partial transpose with respect to the interval $A_2$ for the reduced density matrix $\rho_A$ and the replica limit described as $n_e \rightarrow 1$ is an analytic continuation for even sequences of $n_e$ to $n_e=1$.

For the specific mixed state configuration of adjacent intervals $A_1$ and $A_2$ it could be shown that the quantity $\mathrm{Tr}(\rho_A^{T_2})^{n_e}$ in eq. (\ref {cftneg}) may be expressed as a three point twist correlator on the complex plane which is fixed by the conformal symmetry as 
\begin{equation}\label{threept}
\mathrm{Tr}(\rho_A^{T_2})^{n_e} = 
\langle\mathcal{T}_{n_e}(z_1)\overline{\mathcal{T}}_{n_e}^{2}(z_2)\mathcal{T}_{n_e}(z_3)\rangle = c_n^2 
\frac{C_{\mathcal{T}_n\overline{\mathcal{T}}_n^2\mathcal{T}_n}}{|z_{12}|^{\Delta_{\mathcal{T}^2_{n_e}}}
|z_{23}|^{\Delta_{\mathcal{T}^2_{n_e}}}|z_{13}|^{2\Delta_{\mathcal{T}_{n_e}}-\Delta_{\mathcal{T}^2_{n_e}}}},
\end{equation}
where $|z_{ij}|=|z_{i}-z_{j}|$, and $\Delta_{\tau^2_{n_e}}$ and $\Delta_{\tau_{n_e}}$ are the scaling dimensions of the twist fields $\tau^2_{n_e}$ and $\tau_{n_e}$ respectively. In the large central charge limit this three point twist correlator eq. (\ref{threept}) may be expressed in terms of the lengths of bulk space like geodesics
anchored on the appropriate intervals, through the standard $AdS/CFT$ dictionary as follows \cite{Jain:2017aqk} 
\begin{equation}\label{threeptgeo}
\begin{split}
 \langle\mathcal{T}_{n_e}(z_1)\overline{\mathcal{T}}_{n_e}^{2}(z_2)\mathcal{T}_{n_e}(z_3)\rangle=\exp{\bigg[\frac{-\Delta_{\mathcal{T}_{n_e}}\mathcal{L}_{13}-\Delta_{\mathcal{T}_{\frac{n_e}{2}}}(\mathcal{L}_{12}+\mathcal{L}_{23}-\mathcal{L}_{13})}{R}\bigg]}.
\end{split}
\end{equation}
The holographic entanglement negativity for the mixed state configuration in question may then be expressed
in terms of a specific algebraic sum of the lengths of the bulk space like geodesics as follows
\begin{equation}\label{henconAdS3}
\mathcal{E} = \frac{3}{16G^3_N}(\mathcal{L}_{12}+\mathcal{L}_{23}-\mathcal{L}_{13}), 
\end{equation}
where we have employed the Brown-Henneaux formula ${c}=\frac{3R}{2G^3_N}$  \cite{brown1986}. 
In the context of the $AdS_{d+1}/CFT_d$ scenario the corresponding holographic entanglement negativity for a mixed state of adjacent subsystems $A_1$ and $A_2$ in the $CFT_d$s dual to bulk $AdS_{d+1}$ geometries is given as \cite{Jain:2017xsu}
\begin{equation}\label{HEN CONJ AREA}
\mathcal{E} = \frac{3}{16G^{(d+1)}_N}\big(\mathcal{A}_{1}+\mathcal{A}_{2}-\mathcal{A}_{12}\big),
\end{equation}
where $\mathcal{A}_{i}$'s are the areas of co-dimension two bulk static minimal surfaces anchored on the respective subsystems $A_i$. It may be shown that the above expression reduces to the holographic mutual information ${\cal I}(A_1,A_2)$ between the two intervals (see footnote \ref{fn1}) on utilizing the Ryu-Takayanagi conjecture $\Big(S_{A_i}=\frac{\mathcal{A}_{i}}{4G_N^{(d+1)}}\Big)$ as follows
\begin{equation}\label{hmi}
\mathcal{E} =  \frac{3}{4}(S_{A_1}+S_{A_2}-S_{A_1\cup A_2})=\frac{3}{4}
{\cal I}(A_1, A_2).
\end{equation}

\section{Holographic entanglement negativity for $\mathrm{CFT_3}$ dual to RN-$\mathrm{AdS_4}$}\label{sec3}

As mentioned earlier it is instructive to first examine the application of our holographic entanglement negativity conjecture to a $CFT_3$ with a conserved charge dual to the bulk $AdS_4$ configurations. This exercise will elucidate the non trivial structure of the perturbative expansion for the holographic entanglement negativity for various limits of the charge and the temperature of the dual $CFT_3$. In this context we describe the application of our conjecture
to compute the holographic entanglement negativity for the bipartite mixed state configuration of adjacent subsystems with rectangular strip geometries in the $CFT_3$s dual to bulk non extremal and extremal RN-$AdS_4$ black holes.

\subsection{Area of minimal surface for RN-$AdS_4$ black holes}
We first briefly review the perturbative computation of the area of a co-dimension two bulk static minimal surface anchored on a subsystem of rectangular strip geometry in the dual $CFT_3$ \cite{Fischler:2012ca,Chaturvedi:2016kbk} which will be required for the subsequent calculations. The metric for the RN-$AdS_4$ black hole with a planar horizon ( with the $AdS$ radius $R=1$ ) is given as
\begin{eqnarray}
ds^2 &=& -r^2f(r)dt^2+\frac{1}{r^2f(r)} dr^2+{r^2}(dx^2+dy^2),\label{RNmetric}\\
f(r)&=& 1-\frac{M}{r^3}+\frac{Q^2}{r^4}.\label{lapsefnc}
\end{eqnarray}
The lapse function $f(r)$ vanishes at the horizon $(r=r_h)$ resulting in the following relation between the mass, charge and radius of the horizon as
\begin{equation}\label{MQrel}
 f(r_{h})=0 \Rightarrow M=\frac{r_{h}^4+Q^2}{r_{h}}.
\end{equation}
One may now express the lapse function eq. (\ref{lapsefnc}) in terms of the charge $Q$ and the horizon radius $r_h$ as follows
\begin{equation}\label{RNlapse1}
 f(r)=1-\frac{r_h^3}{r^3}-\frac{Q^2}{r^3r_h}+\frac{Q^2}{r^4}.
\end{equation}
The Hawking temperature for the RN-$AdS_4$ black hole is given as
\begin{equation}\label{RNtemp}
T=\left.\frac{f'(r)}{4\pi}\right|_{r=r_h}=\frac{3 r_{h}}{4\pi}(1-\frac{Q^{2}}{3 r_{h}^{4}}).
\end{equation}

We now proceed to the computation the area of a co-dimension two static minimal surface anchored on a subsystem described by a rectangular strip geometry on the dual $CFT_3$ to the RN-$AdS_4$ black hole. The subsystem $A$ of rectangular strip geometry on the dual $CFT_3$ is specified as follows
\begin{equation}\label{subsys}
x\in[-\frac{l}{2},\frac{l}{2}],~~~~~y\in[-\frac{L}{2},\frac{L}{2}].
\end{equation}
The area ${\cal A}_A$ of the co-dimension two bulk static minimal surface anchored on the subsystem $A$ in the holographic $CFT_3$ may the be expressed as 
\begin{equation}\label{Earea1}
{\cal A}_A=2 L\int^\infty_{r_{c}} \frac{ dr}{\sqrt{f(r)(1-\frac{r_{c}^{4}}{r^{4}}})}.
\end{equation}
The turning point $r_{c}$ of the minimal surface in the bulk, is related to the length of the rectangular strip in the $x$ direction as
\begin{equation}\label{lrel}
  \frac{l}{2}=\int^\infty_{r_{c}} \frac{r_{c}^{2} dr}{r^{4}\sqrt{f(r)(1-\frac{r_{c}^{4}}{r^{4}})}}.
\end{equation}
In order to evaluate these integrals we perform a coordinate transformation from $r$ to $u=\frac{r_c}{r}$ and the eqs. (\ref{RNlapse1}), (\ref{Earea1}) and (\ref{lrel}) may then be expressed as
\begin{eqnarray}
f(u)&=& 1-\frac{{r_h}^3 u^3}{{r_{c}}^3}-\frac{Q^2 u^3}{{r_{c}}^3 {r_h}}+\frac{Q^2 u^4}{{r_{c}}^4},\label{RNlapse2}\\
{\cal A}&=&2Lr_{c}\int _0^1\frac{f(u)^{-\frac{1}{2}} }{u^2\sqrt{1-u^4}}du,\label{Ainu}\\
 l&=&\frac{2}{r_{c}}\int _0^1\frac{u^2 f(u)^{-\frac{1}{2}}}{\sqrt{1-u^4}}du. \label{linu}
\end{eqnarray}
We obtain the area of the minimal surface in question through a perturbative evaluation of the above integrals for different limits of the parameters, charge $Q$ and the temperature $T$ of the $CFT_3$ .

\begin{figure}[H]
\centering
\includegraphics[scale=.3]{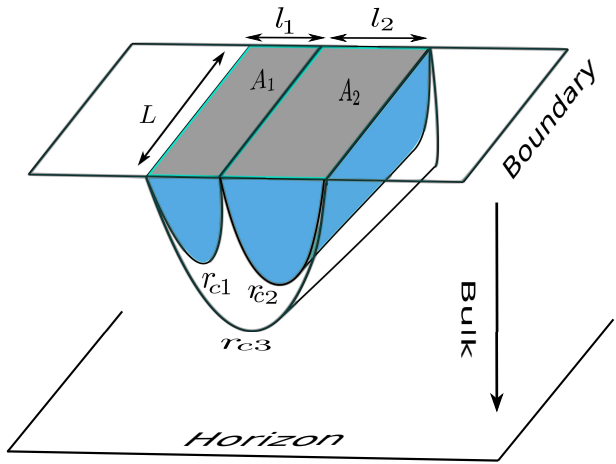}
\caption{Schematic of the bulk static minimal surfaces that are anchored on the subsystems $A_1$, $A_2$ and $A_1 \cup A_2$ on the boundary $CFT_3$ dual to the RN-$AdS_4$ black hole.}\label{fig1}
\end{figure}

In what follows, we compute the holographic entanglement negativity for mixed states of adjacent subsystems described by rectangular strip geometries, in $CFT_3$s dual to RN-$AdS_4$ non-extremal and extremal black holes. 
The  rectangular strip geometries corresponding to these subsystems denoted as $A_1$ and $A_2$, are specified by the coordinates
\begin{eqnarray}
&x\in[-\frac{l_1}{2},\frac{l_1}{2}],~~~~~y\in[-\frac{L}{2},\frac{L}{2}],\label{a1rec}\\
&x\in[-\frac{l_2}{2},\frac{l_2}{2}],~~~~~y\in[-\frac{L}{2},\frac{L}{2}],\label{a2rec}
\end{eqnarray}
respectively, as depicted in Fig. (\ref{fig1}). Note that the areas and the turning points of the corresponding co-dimension two bulk static minimal surfaces anchored on the subsystems $A_1$ and $A_2$, may therefore be obtained from equations (\ref{linu}) and (\ref{Ainu}), by replacing $l$ in eq.(\ref{linu}) by $l_1$ and $l_2$ respectively.

\subsection{Non-extremal RN-$AdS_4$ black holes}

We first consider the finite temperature mixed state configuration of adjacent subsystems with rectangular strip geometries as depicted in Fig. (\ref{fig1}), in the $CFT_3$ dual to a bulk non-extremal RN-$AdS_4$ black hole. To compute the holographic entanglement negativity utilizing our conjecture it is required to evaluate the corresponding areas of the bulk static minimal surfaces perturbatively for various limits of the relevant parameters described above.

\subsubsection{Small charge and low temperature}

The non extremality condition may be obtained in terms of the horizon radius for the bulk RN-$AdS_4$ black hole by setting $T>0$ in eq.(\ref{RNtemp})  as follows
\begin{equation}\label{rhforlowtemp}
r_{h}>\frac{\sqrt{Q}}{{3^{\frac{1}{4}}}}. 
\end{equation}
In the limit of small charge and at low temperatures it may be shown from eq.(\ref{rhforlowtemp}) that  $ r_{h} \ll r_{c} $ and $Q/r_{h}^2\sim 1$. Hence, the function $f(u)^{-\frac{1}{2}}$ is Taylor expanded around $\frac{r_h}{r_c}=0$ to the leading order in $\mathcal{O}[(\frac{r_h}{r_c}u)^3]$ as follows \cite{Chaturvedi:2016kbk}
\begin{equation}\label{lapseRNads1}
f(u)^{-\frac{1}{2}}\approx1+\frac{1+\alpha}{2}\left(\frac{r_h}{r_c}\right)^3 u^3,
\end{equation}
where $f(u)$ is the lapse function eq. (\ref{RNlapse2})for the black hole metric and
$\alpha=\frac{Q^2}{r_h^4}$. Employing the eqs. (\ref{lapseRNads1}), (\ref{linu}) and (\ref{Ainu}), the area of the co-dimension two minimal surface anchored on the subsystem $A$ of rectangular strip geometry, may be expressed as follows \cite {Chaturvedi:2016kbk}
\begin{equation}
{\cal A}_A={\cal A}_A^{div}+{\cal A}_{ A}^{finite},
\end{equation}
where the divergent part ${\cal A}_A^{div}$ and the finite part ${\cal A}_{ A}^{finite}$ of ${\cal A}_A$ are as follows
\begin{eqnarray}
{\cal A}_A^{div}&=&2\Big(\frac{L}{a}\Big),\\
{\cal A}_{ A}^{finite}&= &k_1\frac{L}{l}+k_2r_h^3(1+\alpha)l^2+\mathcal{O}(r_h^4 l^3).
\end{eqnarray}
Here the constants in the above equation are given as 
\begin{eqnarray}
k_1 &=& -\frac{ 4 \pi \Gamma(\frac{3}{4})^2}{\Gamma(\frac{1}{4})^2},\\
k_2 &=&  \frac{\Gamma (\frac{1}{4})^2 }{32\Gamma (\frac{3}{4})^2}.
\end{eqnarray}

The holographic entanglement negativity in the limit of small charge and low temperature, for the mixed state of adjacent subsystems in question may now be obtained from our conjecture using eq. (\ref {HEN CONJ AREA}) as follows 
\begin{equation}\label{ENforSQlowtemp}
\mathcal{E} = \frac{3}{16G_{N}^{3+1}}\Big[\Big(\frac{2L}{a}\Big)+k_1(\frac{L}{l_1}+\frac{L}{l_2}-\frac{L}{l_1+l_2})-2k_2 M L ~l_1 l_2\Big] + \ldots,
\end{equation} 
where the ellipses represent sub leading corrections in this limit. In the above equation note that the second term on the right hand side is identical to the holographic entanglement negativity for the zero temperature mixed state of adjacent subsystems in the $CFT_3$ dual to the bulk pure $AdS_4$ geometry and the second term describes the correction arising from the charge and the temperature.

\subsubsection{Small charge - high temperature}

We next consider the limit of small charge and high temperature given by the conditions $r_h \ll 1$, and $\delta = \frac{Q}{\sqrt{3}r_h^2}\ll 1$ where as earlier $r_h$ represents the horizon radius. In this limit the function $f(u)^{-\frac{1}{2}}$ is Taylor expanded around $\delta=0$ as follows \cite{Chaturvedi:2016kbk}
\begin{equation}\label{lapseRNads2}
f(u)^{-\frac{1}{2}}\approx \frac{1}{\sqrt{1-\frac{{r_h}^3 u^3}{{r_c}^3}}}+
 \frac{3}{2}\left(\frac{r_h}{r_c}\right)^3\frac{\delta ^2  u^3 (1-\frac{{r_h} u}{{r_c}})}
 {(1-\frac{{r_h}^3 u^3}{{r_c}^3})^{3/2})}. 
\end{equation}
Employing the above expression for the lapse function and eqs. (\ref{linu}) and (\ref{Ainu}), the finite part of the area of the co-dimension two bulk minimal surface ${\cal A}_{ A}$ may be expressed as
\begin{equation}\label{shight}
{\cal A}_A^{finite} = Ll r_h^2  + {L} r_h(k_1+\delta^2 k_2)+{L}r_h\epsilon\bigg[k_3+\delta^2 (k_4+k_5 \log \epsilon)\bigg]+O[\epsilon^2],  
\end{equation}
where the constants   $k_1$, $k_2$, $k_3$, $k_4$ and $k_5$ in the above equation are listed in the Appendix (\ref{appen1}) in eqs. (\ref{k1}), (\ref{k2}), (\ref{k3}), (\ref{k4}) and (\ref{k5}) respectively. The parameter $\epsilon$ in the  eq. (\ref{shight}) is given as
\begin{equation}\label{epsilon}
\epsilon = \frac{1}{3}\exp\left(-\sqrt{3}(l r_h - c_1 -c_2 \delta^2)\right),
\end{equation}
where the constants in the above equations are listed in the Appendix (\ref{appen1}) in eq. (\ref{c1}) and (\ref{c2}). The holographic entanglement negativity in this limit for the mixed state in question may then be computed from our conjecture as follows
\begin{eqnarray}\label{ENforSQhightemp}
\mathcal{E} &=&\frac{3 }{16G_{N}^{3+1}}\bigg[\frac{2L}{a}+L r_h\Big\{(k_1+\delta^2 k_2)+
k_3(\epsilon_1+\epsilon_2-\epsilon_{12})\nonumber\\
&&+\delta^2 k_4(\epsilon_1+\epsilon_2-\epsilon_{12})+\delta^2 k_5\big(\log \epsilon_1+\log \epsilon_2-\log \epsilon_{12}\big)\Big\}\bigg]+\ldots,
\end{eqnarray}
where the subscript $i$ in  $\epsilon_i ~(i=1, 2, 12)$ refers to the subsystems $A_1$, $A_2$ and $A_1\cup A_2$ respectively. Interestingly it may be noted that the holographic entanglement negativity
described by the above expression depends only on the length $L$ shared between the adjacent subsystems of rectangular strip geometries (note that this is equivalent to the area of the entangling surface which in the $AdS_4/CFT_3$ scenario reduces to the length). This is unlike the holographic entanglement entropy which scales as the volume (area in the $AdS_4/CFT_3$ scenario) in this limit described in \cite{Chaturvedi:2016kbk}. For the holographic entanglement negativity on the other hand all volume (area in $AdS_4/CFT_3$) dependent thermal contributions cancel leaving a purely area (length in $AdS_4/CFT_3$) dependent expression as expected from quantum information theory. Note that this cancellation is similar to that for the $AdS_3/CFT_2$ case described in \cite{Jain:2017aqk,Chaturvedi:2016rcn} indicating that the elimination of the thermal contribution is possibly a universal feature of the holographic entanglement negativity
in $CFT$s.

\subsubsection{Large charge - high temperature}

For the corresponding large charge and high temperature limit we have the conditions $r_c\sim r_h$ and $u_0=\frac{r_c}{r_h}\sim 1$ as a consequence of the turning point of the co dimension two static minimal surface extending close to the horizon in the bulk.  In this case the Taylor expansion for the function $f(u)^{-\frac{1}{2}}$ around 
$u_0$  is given as \cite{Chaturvedi:2016kbk}
\begin{equation}\label{lapseRNads4}
 f(u)\approx\Big(3-\frac{Q^2}{r_h^4}\Big)\Big(1-\frac{r_h}{r_c}u\Big).
\end{equation}
The finite part of the area of the bulk  static minimal surface in this case may be expressed as
\begin{equation}\label{EEforLQLTemp}
 {\cal A}_A^{finite}= Llr_h^2+\frac{Lr_h}{2\sqrt{\delta}}\bigg[K_1'+K_2'\epsilon+\mathcal{O}(\epsilon^2)\bigg],
\end{equation}
where $\epsilon$ is given by the eq. (\ref{epsilon}), and the  constants $K'_1$ and $K'_2$ are  listed in the Appendix (\ref{appen2}) in the eqs. (\ref{K'1}) and (\ref{K'2}).

The holographic entanglement negativity for the mixed state configuration of adjacent subsystems in this limit may then be  obtained from our conjecture as follows
\begin{equation}\label{ENforLQhightemp}
\mathcal{E} =\frac{3}{8G_{N}^{3+1}}
\Bigg[\Big(\frac{L}{a}\Big)+\frac{L r_h}{\sqrt{\delta}}\Big\{K_1'+K_2'(\epsilon_1+\epsilon_2-\epsilon_{12})\Big\} \Bigg]+\ldots,
\end{equation}
where the subscript $i$ in  $\epsilon_i ~(i=1, 2, 12)$ refers to the subsystems $A_1$, $A_2$ and $A_1\cup A_2$ respectively. Note that as earlier the volume (area in $AdS_4/CFT_3$) dependent thermal terms cancel and the entanglement negativity scales as the area (length in $AdS_4/CFT_3$) of the
entangling surface as expected from quantum information theory. Once again the elimination of thermal contribution is similar to the corresponding $AdS_3/CFT_2$ case indicating that it is an universal feature for the holographic entanglement negativity for $CFT$s.

\subsection{Extremal RN-$AdS_4$ black holes}

Having described the holographic entanglement negativity for the required mixed state  in the $CFT_3$ with a conserved charge, dual to the bulk non extremal RN-$AdS_4$ black hole, we now turn our attention to the corresponding extremal case. To this end we consider the zero temperature mixed state configuration of adjacent subsystems with rectangular strip geometries in the $CFT_3$ dual to the bulk extremal  RN-$AdS_4$ black hole . Here we describe the computation of the holographic entanglement negativity from our conjecture, perturbatively in the limits of small and large charge.

\subsubsection{Small charge - extremal}
 
In the limit of small charge, the function $f(u)^{-\frac{1}{2}}$ may be Taylor expanded around $\frac{r_h}{r_c}=0$  to the leading order in $\mathcal{O}[(\frac{r_h}{r_c}u)^3]$  as follows \cite{Chaturvedi:2016kbk}
\begin{equation}\label{fapproxExtrm} 
f(u)^{-\frac{1}{2}}\approx1+2\frac{r_{h}^3}{r_c^3}u^3, 
\end{equation}
Now employing eqs. (\ref{fapproxExtrm}), \eqref{linu} and \eqref{Ainu} it is possible to express the finite part of the area of the bulk co- dimension two static minimal surface anchored on the subsystem $A$ as 
\begin{equation}\label{EEareaExtrm1} 
 {\cal A}_{ A}^{finite}= k_1\frac{L}{l}+k_2 r_h^3Ll^2+\mathcal{O}(r_h^4l^3),
\end{equation}
where the constants are given as follows
\begin{eqnarray*}
 k_1&=&-\frac{4 \pi \Gamma(\frac{3}{4})^2}{4\Gamma(\frac{1}{4})^2},\\
k_2&=& \frac{4\Gamma (\frac{1}{4})^2 }{32\Gamma (\frac{3}{4})^2}.
\end{eqnarray*}
The holographic entanglement negativity of the mixed state in question may then be obtained from our conjecture 
as follows
\begin{equation}\label{ENforRNSCext}
\mathcal{E} = \frac{3}{16 G_{N}^{3+1}}\Big[\Big(\frac{2L}{a}\Big)+k_1(\frac{L}{l_1}+\frac{L}{l_2}-\frac{L}{l_1+l_2})-2k_2r_h^{3} L l_1 l_2\Big]+\ldots.
\end{equation}
The first two terms in the above equation describe the holographic entanglement negativity for the zero temperature mixed state of adjacent subsystems in the $CFT_3$ dual to the bulk pure $AdS_4$ geometry and the third term describes the correction arising from the conserved charge of the extremal RN-$AdS_4$ black hole.


\subsubsection{Large charge }
As explained in the earlier sections, in the limit of large charge for the bulk extremal RN-$AdS_4$ black hole we have the ratio $u_0=\frac{r_c}{r_h}\sim 1$. In this case the function $f(u)^{-\frac{1}{2}}$ may be Taylor expanded around $u=u_0$ as follows \cite{Chaturvedi:2016kbk}
\begin{equation}\label{lapseRNads3}
 f(u)\approx 6\Big(1-\frac{r_{h}}{r_{c}}u\Big)^2. 
\end{equation}
Now utilizing the eqs. \eqref{lapseRNads3}, \eqref{linu} and \eqref{Ainu}, the finite part of the area of the co dimension two static minimal  surface anchored on the subsystem with rectangular strip geometry in the $CFT_3$ dual to the bulk extremal RN-$AdS_4$ black hole in the large charge limit, may then be expressed as
entropy \cite{Chaturvedi:2016kbk}
\begin{equation}\label{EEforLQExtr}
 {\cal A}_A^{finite}=Ll r_h^2 + {Lr_h}  \Big(K_1+K_2\sqrt{\epsilon}+K_3\epsilon+\mathcal{O}(\epsilon^\frac{3}{2})\Big),
\end{equation}
where the constants $K_1$, $K_2$ and $K_3$  appearing in the above expression are listed in the Appendix (\ref{appen3}) in the eqs. (\ref{K1}), (\ref{K2}) and (\ref{K3}).

The holographic entanglement negativity for the mixed state of adjacent subsystems with rectangular strip geometries in the dual $CFT_3$ may then be obtained from our conjecture in the large charge limit as follows
\begin{equation}\label{ENforLQzerotemp}
\mathcal{E} =\frac{3}{16G_{N}^{3+1}}
\bigg[\Big(\frac{2L}{a}\Big)+L r_h\Big\{K_1+K_2(\sqrt{\epsilon_1}+\sqrt{\epsilon_2}-\sqrt{\epsilon_{12}}) + K_3(\epsilon_1+\epsilon_2-\epsilon_{12})\Big\} \bigg]+\ldots,
\end{equation}
where the subscripts in  $\epsilon_i ~(i=1, 2, 12)$ refer to the subsystems $A_1$, $A_2$ and $A_1\cup A_2$ respectively.\\
Interestingly even for the extremal case in the large charge limit we once again observe that the holographic entanglement negativity for the zero temperature mixed state, following from our conjecture is purely dependent on the area (length in the $AdS_4/CFT_3$ scenario). As earlier for the non extremal case the volume (area for the $AdS_4/CFT_3$ case) dependent contributions arising from the counting entropy of the degenerate $CFT_3$ vacuum in this case cancel leaving a purely area dependent expression as in specific earlier cases described in previous sections.

The above results for the holographic entanglement negativity of the mixed state
configurations of adjacent subsystems in the dual $CFT_3$ for various limits of the relevant parameters, conform to quantum information expectations. It is observed that in the  large charge and/or large temperature regimes where the holographic entanglement entropy is dominated by volume dependent thermal contributions, the corresponding entanglement negativity depends purely on the area of the entangling surface in the $CFT$. This arises from the exact cancellation of the volume dependent thermal terms between the appropriate combinations of the contributions from the adjacent subsystems. As remarked earlier this cancellation is similar to that observed for certain $AdS_3/CFT_2$ examples and seems to be a universal feature of $CFT$s. Naturally, these results constitute strong consistency checks substantiating the higher dimensional extension of our conjecture.

Having obtained the holographic entanglement negativity for the mixed state of adjacent subsystems with rectangular strip geometries in the $AdS_4/CFT_3$ scenario and illustrating the non trivial structure of the limits associated with the perturbative expansion we now turn our attention to the generic $AdS_{d+1}/CFT_d$ scenario in the next section.

\section{Holographic entanglement negativity for $\mathrm{CFT_d}$ dual to RN-$\mathrm{AdS_{d+1}}$}\label{sec4}

In this section following our earlier analysis for the holographic entanglement negativity of mixed states of adjacent subsystems in the $AdS_4/CFT_3$ scenario, we now proceed to examine the corresponding case for the $AdS_{d+1}/CFT_d$ scenario. As earlier this case also involves a perturbative evaluation of the areas of the corresponding bulk static minimal surfaces anchored on the respective subsystems in various limits of the appropriate parameters of the 
RN-$AdS_{d+1}$ black hole which in this case are the temperature $T$ and the chemical potential $\mu$ conjugate to the charge $Q$. Note however that for the $AdS_{d+1}/CFT_d$ scenario it is convenient to describe the holographic entanglement negativity in terms of an {\it effective temperature} $T_{\mathrm{eff}}$ and another parameter $\varepsilon$ which is a function of the temperature and the chemical potential, describing the total energy of the dual $CFT_d$. The parameter $\varepsilon$  is therefore  related to the expectation value of the $T_{00}$ component of the energy momentum tensor \cite{Kundu:2016dyk}.

\subsection{Area of minimal surfaces in RN-$\mathrm{AdS_{d+1}}$}

The metric for the ${RN-AdS_{d+1}}$ ($d\geq3$) black hole with the $AdS$ length scale $R=1$  is given as 
\begin{equation}\label{henvacuum}
\begin{aligned}
 ds^2 =& \frac{1}{z^2} \left(- f(z) dt^2 + \frac{dz^2}{f(z)} + d\vec{x}^2 \right),\\
 f(z) = &1- M z^d + \frac{(d-2)Q^2}{(d-1) } z^{2(d-1)},\\
 A_t =& Q (z_H^{d-2} - z^{d-2}),
\end{aligned}
\end{equation}
where $Q$ and $M$ are the mass and charge of the black hole respectively.  The location of the horizon $z_H$ is given by the smallest real root of the lapse function $f(z)=0$. The corresponding chemical potential $\mu$ conjugate to the charge $Q$ is defined as follows
\begin{equation}\label{muforddim}
\mu \equiv \lim_{z\to 0} A_t(z) = {Q} z_H^{d-2}, 
\end{equation}
and the Hawking temperature is 
\begin{equation}\label{Hawktempforddim}
T=-\frac{1}{4\pi}\frac{d}{dz}f(z)\bigg|_{z_H}=\frac{d}{4 \pi z_H}\left(1-\frac{(d-2)^2 Q^2 z_H^{2(d-1)}}{d(d-1)}\right). 
\end{equation}
The lapse function, chemical potential and the temperature may now be expressed as follows
\begin{eqnarray}
f(z)& = &1- \varepsilon \left(\frac{z}{z_H}\right)^d + 
\left(\varepsilon-1\right) \left(\frac{z}{z_H}\right)^{2(d-1)}, \label{RNlapseddim2} \\ 
\mu&=&\frac{1}{z_H}\sqrt{\frac{ (d-1)}{(d-2)}(\varepsilon-1)}, \label{muforddim2} \\
T&=&\frac{ 2(d-1)-(d-2)\varepsilon}{4\pi z_H}. \label{Tforddim2}
\end{eqnarray}
Here $\varepsilon$ is a dimensionless quantity with limits $1\ge\varepsilon\ge \frac{2(d-1)}{d-2}$, that describes the energy of the system \cite{Kundu:2016dyk}, as follows
\begin{equation}\label{epsilonforddim}
\varepsilon(T,\mu)=b_0-\frac{2n}{1+\sqrt{1+\frac{d^2}{2\pi^2 b_0b_1}\left(\frac{\mu^2}{T^2}\right)}},
\end{equation}
where the constants $b_0$ and $b_1$  are given as 
\begin{equation}\label{abforddim}
b_0=\frac{2(d-1)}{d-2}\ , \qquad b_1=\frac{d}{d-2}.
\end{equation}
The effective temperature $T_{eff}$ describing the number of microstates for a given temperature and chemical potential may be defined as \cite{Kundu:2016dyk}
\begin{equation}\label{Teffforddim}
T_{\mathrm{eff}}(T,\mu) \equiv \frac{d}{4\pi z_H}=
\frac{T}{2}\left[1+\sqrt{1+\frac{d^2}{2\pi^2 b_0 b_1}\left(\frac{\mu^2}{T^2}\right)}\right].
\end{equation}

We now proceed to compute the area of a co dimension two bulk static minimal surface anchored on a subsystem with rectangular strip geometry in the dual $CFT_d$. The strip geometry of the subsystem in question may then be specified as follows
\begin{equation}\label{stripforddim}
x\equiv x^1 \in \left[-\frac{l}{2},
\frac{l}{2}\right],~  x^i\in \left[-\frac{L}{2},\frac{L}{2}\right],\qquad i=2,...,d-2,
\end{equation}
where $L \rightarrow \infty$. The area $\mathcal{A}$ of the co-dimension two bulk extremal surface  anchored on the subsystem in  the boundary may be expressed as
\begin{equation}\label{areaforddim}
 \mathcal{A}=2L^{d-2}z_*^{d-1}\int_{0}^{l/2}\frac{dx}{z(x)^{2(d-1)}}=
2L^{d-2}z_*^{d-1}\int_{a}^{z_*}\frac{dz}{z^{d-1}\sqrt{f(z)[z_*^{2(d-1)}-z^{2(d-1)}]}},
\end{equation}
where $a$ is the UV cut off of the $CFT_d$. The turning point $z_*$ of the extremal surface in the bulk is related to $l$, the length of the strip in the $x^1$ direction  as
\begin{equation}\label{z*forddim}
\frac{l}{2}=\int_0^{z_*}\frac{dz}{\sqrt{f(z)[(z_*/z)^{2(d-1)}-1]}}.
\end{equation}
The authors in \cite{Kundu:2016dyk} demonstrated that the above integral may be expressed as a double sum as 
\begin{align}
l=\frac{z_*}{d-1}\sum_{n=0}^\infty \sum_{k=0}^n \frac{\Gamma\left[\frac{1}{2}+n\right]\Gamma \left[\frac{d (n+k+1)-2k}{2 (d-1)}\right]\varepsilon^{n-k} (1-\varepsilon)^k}{
\Gamma[1+n-k]\Gamma[k+1]\Gamma \left[\frac{d (n+k+2)-2k-1}{2 (d-1)}\right]}  \left(\frac{z_*}{z_H}\right)^{n d +k(d-2)} \ .\label{eerc}
\end{align}
The area of the static minimal surface may also be expressed as a double sum as follows \cite{Kundu:2016dyk}
\begin{align}\label{aren}
\mathcal{A}&= \frac{2}{d-2}\left(\frac{L}{a}\right)^{d-2}+2 \frac{L^{d-2}}{z_*^{d-2}} \left[\frac{\sqrt{\pi } \Gamma \left(-\frac{d-2}{2
(d-1)}\right)}{2 (d-1) \Gamma \left(\frac{1}{2 (d-1)}\right)}\right]\\
&+  \frac{L^{d-2}}{(d-1)z_*^{d-2}} \left[\sum_{n=1}^\infty \sum_{k=0}^n\frac{\Gamma\left[\frac{1}{2}+n\right]\Gamma \left[\frac{d (n+k-1)-2k+2}{2
(d-1)}\right]\varepsilon^{n-k} (1-\varepsilon)^k}{ \Gamma[1+n-k]\Gamma[k+1]\Gamma \left[\frac{d (n+k)-2k+1}{2 (d-1)}\right]}   \left(\frac{z_*}{z_H}\right)^{n d
+k(d-2)}\right].\nonumber
\end{align}
The area and the turning point of the corresponding static minimal surface are expressed in terms of the specified parameters $T_{eff}$ and $\varepsilon$  as a perturbation expansion, for various limits of the chemical potential $\mu$ and the temperature $T$ of the dual $CFT_d$. We now proceed to describe these evaluations and utilize them to obtain the
holographic entanglement negativity for the mixed states in question, from our conjecture.

In what follows, we compute the holographic entanglement negativity for mixed states of adjacent subsystems described by rectangular strip geometries in $CFT_d$s dual to RN-$AdS_{d+1}$ non-extremal and extremal black holes. The  rectangular strip geometries corresponding to these subsystems denoted as $A_1$ and $A_2$, are specified by the coordinates
\begin{eqnarray}
&x^1\in[-\frac{l_1}{2},\frac{l_1}{2}],~~~~~x^i\in[-\frac{L}{2},\frac{L}{2}],\label{a1recd}\\
&x^1\in[-\frac{l_2}{2},\frac{l_2}{2}],~~~~~x^i\in[-\frac{L}{2},\frac{L}{2}],\label{a2recd}
\end{eqnarray}
respectively, as depicted in Fig. (\ref{fig1}) (with $L$ now denoting the length of the strip in the remaining $(d-2)$ directions). Note that the areas and the turning points of the corresponding co-dimension two bulk static minimal surfaces anchored on the subsystems $A_1$ and $A_2$, may therefore be obtained from equations (\ref{areaforddim}) and (\ref{z*forddim}), by replacing $l$ in eq.(\ref{z*forddim}) by $l_1$ and $l_2$ respectively.

\subsection{Non-extremal RN-$\mathrm{AdS_{d+1}}$}

We first consider the non-extremal ${RN-AdS_{d+1}}$ black holes and compute the holographic entanglement negativity for the finite temperature mixed state of adjacent subsystems described by rectangular strip geometries in the dual $CFT_d$ fr various limits of the chemical potential
$\mu$ and the temperature $T$.

\subsubsection{Small chemical potential - low temperature}

The limit of small chemical potential and low temperature is defined by the conditions $Tl\ll1$ and $\mu l\ll1$. Notice that  apart from the chemical potential $\mu$ and the temperature $T$, the area of the static minimal surface depends on the length of the rectangular strip along the $x^1$ direction provided we keep the lengths in all the other $x^i$ direction to be the constant $L$.  Hence, the limit of small chemical potential and low temperature has to be fixed by specifying another condition which is chosen to be $T\ll\mu$ or $T\gg\mu$ as described in \cite{Kundu:2016dyk}. Below we compute the holographic entanglement negativity of the required mixed state for both of the above mentioned limits.
\smallskip \\

\noindent{\boldmath$ (i) ~~T l\ll \mu l\ll 1$ }
\vspace{0.25cm}

\noindent
We first consider the limit defined by the conditions $Tl\ll1$ , $\mu l\ll1$ and $T\ll \mu$ which may be re-casted as $T l\ll \mu l\ll 1$. In the limit  $T\ll \mu$,  the parameters $T_{\mathrm{eff}}(T,\mu)$ and $\varepsilon(T,\mu)$ described by eq.(\ref{Teffforddim}) and eq.(\ref{epsilonforddim}) may be approximated by Taylor expanding them around $\frac{T}{\mu}=0 $ to the leading order as follows\cite{Kundu:2016dyk} 
\begin{align}
T_{\mathrm{eff}}&\approx \frac{1}{2}\left(\frac{\mu d}{\pi\sqrt{2b_0b_1}}+T\right)\label{ttllm},\\
\varepsilon &\approx b_0-\frac{2n\pi \sqrt{2b_0b_1}}{d}\left(\frac{T}{\mu}
\right)\label{vtllm}.
\end{align}
Now from the other two conditions $Tl\ll1$  and $\mu l\ll1$  it may be shown that the turning point of the static minimal surface is far from the horizon i.e $z_*\ll z_H$. Hence the  expression for the turning point may be obtained by expanding eq.(\ref{eerc}) to the leading order in  $(\frac{l}{z_H})^d$ as
\begin{align}
z_*=\frac{l~\Gamma\left[\frac{1}{2(d-1)}\right]}{2 \sqrt{\pi}\Gamma\left[\frac{d}{2(d-1)}\right]} \left[1- \frac{1}{2(d+1)}\frac{2^{\frac{1}{d-1}-d} \Gamma
	\left(1+\frac{1}{2 (d-1)}\right) \Gamma \left(\frac{1}{2 (d-1)}\right)^{d+1}}{\pi^{\frac{d+1}{2}}  \Gamma
	\left(\frac{1}{2}+\frac{1}{d-1}\right)\Gamma\left(\frac{d}{2(d-1)}\right)^d}\varepsilon\left(\frac{l}{z_H}\right)^d\right.\nonumber\\
\left.+\mathcal{O}\left(\frac{l}{z_H}\right)^{2(d-1)}\right]\ \label{turnsq}.
\end{align}
Similarly, the area of the minimal surface may be obtained by  eq.(\ref{aren}) to the leading order in  $(\frac{l}{z_H})^d$ and is re-expressed in terms of $T_{\mathrm{eff}}$ and $\varepsilon$ as  follows \cite{Kundu:2016dyk} 
\begin{equation}\label{areasq}
{\cal A}_A = \bigg[\frac{2}{d-2}\left(\frac{L}{a}\right)^{d-2} + 
\mathcal{S}_0 \left(\frac{L}{l}\right)^{d-2} + \varepsilon\mathcal{S}_0\mathcal{S}_1
\left( \frac{4\pi T_{\mathrm{eff}}}{d}\right)^d
 L^{d-2} l^2
\bigg]+\mathcal{O}\Big( T_{\mathrm{eff}} l\Big)^{2(d-1)}.
\end{equation}
The holographic entanglement negativity for the mixed state in question, is then given as
\begin{align}
\mathcal{E} = \frac{3}{16G_N^{d+1}}\bigg[\frac{2}{d-2}\left(\frac{L}{a}\right)^{d-2} + 
\mathcal{S}_0 L^{d-2}\left(\frac{1}{l_1^{d-2}} +  \frac{1}{l_2^{d-2}} - \frac{1}{(l_1+l_2)^{d-2}}\right)\nonumber \\ - 
\varepsilon\mathcal{S}_0\mathcal{S}_1
\left(\frac{4\pi T_{\mathrm{eff}}}{d} \right)^d L^{d-2} 2l_1l_2\   \bigg]+\ldots.\label{tllm}
\end{align}
In the above expression for the holographic entanglement negativity for the finite temperature mixed state the first two terms are identical to those in the holographic negativity for the zero temperature mixed state of adjacent subsystems in the $CFT_d$ which is dual to the bulk pure $AdS_{d+1}$ geometry. The other term describes the correction due to the chemical potential and the temperature of the black hole.
\smallskip \\
\newpage
\noindent{\boldmath$ (ii) ~~\mu l\ll T l\ll 1$ }
\vspace{0.25cm}

\noindent
We consider the limit  $Tl\ll1$ , $\mu l\ll1$ and $T\gg \mu$ which may be recast as $\mu l\ll T l\ll 1$. In this limit the parameters $T_{\mathrm{eff}}(T,\mu)$ and $\varepsilon(T,\mu)$ described by eq.(\ref{Teffforddim}) and eq.(\ref{epsilonforddim}) may be Taylor expanded around $\frac{\mu}{T}=0$ to the leading order as follows \cite{Kundu:2016dyk}
\begin{align}
&T_{\mathrm{eff}}=T\left[1+\frac{d(d-2)^2}{16\pi^2(d-1)}
\left(\frac{\mu}{T}\right)^2+\mathcal{O}\left(\frac{\mu}{T}\right)^4\right]\ ,\label{Teffforddim2}\\
&\varepsilon=1+\frac{d^2(d-2)}{16\pi^2(d-1)}\left(\frac{\mu}{T}\right)^2+
\mathcal{O}\left(\frac{\mu}{T}\right)^4\ \label{epsilonforddim2}.
\end{align}
Once again from the conditions $Tl\ll1$ and $\mu l\ll1$ it is clear that $z_*\ll z_H$. The expressions for the turning point may be obtained by expanding eq.(\ref{eerc}) to the leading order in  $(\frac{l}{z_H})^d$ and is same as the one given in eq.(\ref{turnsq}). The area of the co dimension two minimal surface anchored on the subsystem $A$ of rectangular strip geometry in the dual $CFT_d$, is once again determined by expanding eq.(\ref{aren}) to the leading order in  $(\frac{l}{z_H})^d$  as \cite{Kundu:2016dyk}
\begin{equation}\label{EEforddimLQlowtemp}
~~{\cal A}_A = \Big[\frac{2}{d-2}\left(\frac{L}{a}\right)^{d-2} + 
\mathcal{S}_0 \left(\frac{L}{l}\right)^{d-2} + \varepsilon\mathcal{S}_0\mathcal{S}_1
\left( \frac{4\pi T_{\mathrm{eff}}}{d}\right)^d
 L^{d-2} l^2
\Big]+\mathcal{O}\Big( T_{\mathrm{eff}} l\Big)^{2(d-1)},
\end{equation} 
where the numerical constants $\mathcal{S}_{0}$ and $\mathcal{S}_{1}$ are listed in the Appendix (\ref{B1}) in the eqs. (\ref{s0}) and (\ref{s1}) respectively. Note that although  the area of the static minimal surface given in eq.(\ref{tllm}) and eq.(\ref{HENmllt}) have identical forms for both $Tl \ll \mu l \ll 1$  and $\mu l\ll T l\ll 1$, the expressions for the effective temperature $T_{\mathrm{eff}}$ and the parameter $\varepsilon$, for this case are  distinct as given by equations   (\ref{Teffforddim2})  and (\ref{epsilonforddim2}).

The holographic entanglement negativity for the required finite temperature mixed state of adjacent subsystems with rectangular strip geometries, in the limit of small charge and low temperature may then be given from our conjecture as follows
\begin{align}
\mathcal{E} = \frac{3}{16G_N^{d+1}}\Big[\frac{2}{d-2}\left(\frac{L}{a}\right)^{d-2} + 
\mathcal{S}_0 L^{d-2}\left(\frac{1}{l_1^{d-2}} +  \frac{1}{l_2^{d-2}} - \frac{1}{(l_1+l_2)^{d-2}}\right)\nonumber \\ - 
\varepsilon\mathcal{S}_0\mathcal{S}_1
\left(\frac{4\pi T_{\mathrm{eff}}}{d} \right)^d L^{d-2} 2l_1l_2\   \Big]+\ldots.\label{HENmllt}
\end{align}
Once again it may observed that the holographic entanglement negativity for the finite temperature mixed state contains three terms. The first two terms are identical to the holographic negativity for the zero temperature mixed state of adjacent subsystems in the $CFT_d$ which is dual to the bulk pure $AdS_{d+1}$ geometry. The third term describes the correction due to the chemical potential and the temperature of the black hole.

\subsubsection{Small chemical potential - high temperature}

Having computed the holographic entanglement for the mixed state in question in the limit of small chemical and low temperature, we now proceed to obtain the same in the limit of small chemical potential and high temperature. This limit is defined by the conditions $\mu\ll T$ and $Tl\gg1$ as described in \cite{Kundu:2016dyk}. 
 As explained in the previous subsection for $\mu\ll T$, the parameter $T_{\mathrm{eff}}$ and $\varepsilon$ may be approximated by Taylor expanding them around $\frac{\mu}{T}=0$, to the leading order in $\frac{\mu}{T}$ as given by eq.(\ref{Teffforddim2}) and eq.(\ref{epsilonforddim2}). However, in contrast to the previous case, the other condition $Tl\gg1$ implies that the turning point of the  minimal surface is close to the horizon i.e $z_*\sim z_H$. Hence, the area of the static minimal surface may be obtained perturbatively from eqs. (\ref{aren}) by expanding it around $\frac{z_*}{z_H}=1$ as follows 

\begin{equation}\label{EEforddimSQhightemp}
{\cal A}_A =\Big[\frac{2}{d-2}\left(\frac{L}{a}\right)^{d-2} + 
V\left( \frac{4\pi T_{\mathrm{eff}}}{d}\right)^{d-1}
+ L^{d-2}\left( \frac{4\pi T_{\mathrm{eff}}}{d}\right)^{d-2}\gamma_{d}\left(\frac{\mu}{T}\right)\Big], 
\end{equation} 
where $V=L^{d-2}l$ is the volume of the strip, and the function $\gamma_{d}\Big(\frac{\mu}{T}\Big)$  in the above expression is perturbative in $\frac{\mu}{T}$ as given in the Appendix (\ref{B2}) in the eq. (\ref{gammad}).

The holographic entanglement negativity for the mixed state in question may then be obtained from our conjecture as follows
\begin{equation}\label{ENforddim3}
\mathcal{E} = \frac{3}{16G_N^{d+1}}\Big[\frac{2}{d-2}\left(\frac{L}{a}\right)^{d-2}+
L^{d-2}\left( \frac{4\pi T_{\mathrm{eff}}}{d}\right)^{d-2}\gamma_{d}\left(\frac{\mu}{T}\right)\Big].
\end{equation}
Note that in the above equation the leading contribution to the holographic entanglement negativity for the mixed state in this limit is purely dependent on the area of the entangling surface between the adjacent subsystems with rectangular strip geometries. As earlier for the $AdS_4/CFT_3$ case the volume dependent thermal contributions cancel leaving an expression that is purely area dependent. This once again conforms to the standard quantum information expectations for the negativity and serves as a strong consistency check for our conjecture.


\subsubsection{Large chemical potential - low temperature}

The limit of large chemical potential and low temperature is described by the conditions $\mu l\gg1$ and $T\ll \mu$.
As explained earlier utilizing the condition $T\ll \mu$, the parameters $T_{\mathrm{eff}}(T,\mu)$ and $\varepsilon(T,\mu)$ may be approximated by Taylor expanding them around $\frac{T}{\mu}=0$ as given by eq.(\ref{ttllm}) and eq.(\ref{vtllm}) respectively.
Employing the other condition $\mu l\gg1$ implies that the turning point of the  minimal surface is close to the horizon i.e $z_*\sim z_H$. Hence, the area of the static minimal surface may once again be obtained perturbatively from eqs. (\ref{aren}) by expanding it around $\frac{z_*}{z_H}=1$ as follows  \cite{Kundu:2016dyk}
\begin{equation}\label{EEforddimLQhightemp}
{\cal A}_A = \Big[\frac{2}{d-2}\left(\frac{L}{a}\right)^{d-2} + 
V\left( \frac{4\pi T_{\mathrm{eff}}}{d}\right)^{d-1}
+ L^{d-2}\left( \frac{4\pi T_{\mathrm{eff}}}{d}\right)^{d-2}\Big(N_0 + N_1(b_0-\varepsilon)\Big)+\mathcal{O}\left(\frac{T}{\mu}
\right)\Big], 
\end{equation}
where $V=L^{d-2}l$ is the volume of the strip, and the numerical constant in the above expression $ N_{0}$ and $N_{1}$ are listed in the Appendix (\ref{B3}) in the eqs. (\ref{N0}) and (\ref{N1}) respectively.

The holographic entanglement negativity for the finite temperature mixed state of adjacent subsystem in question may then be obtained from our conjecture as follows
\begin{equation}\label{ENforddim5}
\mathcal{E} = \frac{3}{16G_N^{d+1}}\Big[\frac{2}{d-2}\left(\frac{L}{a}\right)^{d-2}
+ L^{d-2}\left( \frac{4\pi T_{\mathrm{eff}}}{d}\right)^{d-2}\Big(N_0 + N_1(b_0-\varepsilon)\Big)\Big] + \ldots.
\end{equation}
We observe from the above expression that in the limit of large chemical potential and high temperature,  the holographic entanglement negativity obtained from our conjecture is purely dependent on the area of the entangling surface between the
adjacent subsystems with rectangular strip geometries. As earlier this indicates the cancellation of the volume dependent thermal contributions conforming to the usual quantum information theory expectations and constitutes yet another fairly strong consistency check for our conjecture.

\subsection{Extremal RN-$\mathrm{AdS_{d+1}}$}
Having obtained the holographic entanglement negativity for the finite temperature mixed state in the $CFT_d$ dual to the bulk non extremal RN-$AdS_{d+1}$ black hole, we now turn our attention to the zero temperature mixed state dual to
a bulk extremal RN-$AdS_{d+1}$ black hole. The relevant parameters in this case are given as \cite{Kundu:2016dyk}
\begin{align}
Q^2=&d(d-1)L^2/(d-2)^2z_H^{2(d-1)},\label{Qext}\\
\varepsilon=&b_1,\\
\mu=&\frac{1}{z_H}\sqrt{\frac{b_0b_1}{2}}=\frac{1}{z_H}\sqrt{\frac{d (d-1)}{(d-2)^2}},\\
T_{\mathrm{eff}}=&\frac{\mu d}{2\pi\sqrt{2 b_0b_1}}.
\end{align}
Here $Q$ represents the charge of the extremal RN-$AdS_{d+1}$ black hole and $T_{eff}$ is the effective temperature as earlier. Using the parameters as given above we now proceed to obtain the area of a co dimension two bulk minimal surface
anchored on a subsystem with rectangular strip geometry in a perturbative expansion for various limits of the charge $Q$.
The area expression may then be utilized to obtain the holographic entanglement negativity for the mixed state in question from our conjecture.

\subsubsection{Small chemical potential}

Note that in the small chemical potential limit  defined by the condition $\mu l\ll1$ the turning point of the static minimal surface is far away from the horizon $z_*\ll z_H$ . Hence  the equation (\ref{z*forddim}) may be solved for $z_*$
and at leading order in $(l/z_H)^d$ which once again leads to eq.(\ref{turnsq})\cite{Kundu:2016dyk}. The area of the minimal surface anchored on the subsystem-$A$ of rectangular strip geometry may then obtained perturbatively expanding eq.(\ref{aren}) to the leading order in $(l/z_H)^d$ . Upon re-expressing the area of the static minimal surface  in terms of $\mu$, it is possible to show that \cite{Kundu:2016dyk}
\begin{equation}\label{EEforddimLQExtr}
{\cal A}_A =\Big[\frac{2}{d-2}\left(\frac{L}{a}\right)^{d-2} + 
\mathcal{S}_0 \left(\frac{L}{l}\right)^{d-2} + \mathcal{S}_0\mathcal{S}_1 \frac{2 (d-1)}{d-2}
\left(\frac{(d-2)\mu}{\sqrt{d(d-1)}} \right)^d L^{d-2} l^2
+\mathcal{O}[(\mu l)^{2(d-1)}]\Big],
\end{equation} 
where the constants $\mathcal{S}_0,~\mathcal{S}_1$ are identical to earlier cases and given in the Appendix. The holographic entanglement negativity  for the mixed state of the adjacent subsystem of rectangular strip geometries, in the small charge limit may then be obtained utilizing our conjecture as follows
\begin{equation}\label{ENforddim1}
\begin{aligned}
\mathcal{E} =  \frac{3}{16G_N^{d+1}}\Big[\frac{2}{d-2}\left(\frac{L}{a}\right)^{d-2} + 
\mathcal{S}_0 L^{d-2}\left(\frac{1}{l_1^{d-2}} +  \frac{1}{l_2^{d-2}} - \frac{1}{(l_1+l_2)^{d-2}}\right) \\ - 
\mathcal{S}_0\mathcal{S}_1 \frac{2 (d-1)}{d-2}
\left(\frac{(d-2)\mu}{\sqrt{d(d-1)}} \right)^d L^{d-2} 2l_1l_2\ \Big]+\ldots.
\end{aligned}
\end{equation}
Observe that the first two terms in the above expression correspond to the holographic entanglement negativity for the zero temperature mixed state of adjacent subsystems with rectangular strip geometries in the $CFT_d$  dual to the bulk pure $AdS_{d+1}$ geometry. The other term along with the sub leading higher order terms describe the correction due to the chemical potential of the $CFT_d$ .

\subsubsection{Large chemical potential}

For the case of extremal  ${RN-AdS_{d+1}}$ black hole , the limit of the  large chemical potential is specified  by the condition $\mu l\gg1$. Hence, it may be observed from eq.(\ref{Qext}) that the horizon radius is large and  the turning point of the static minimal surface is therefore  close to the horizon i.e $z_*\to z_H$ . The area of the static minimal surface anchored on  the subsystem-$A$ of rectangular strip geometry may be obtained by evaluating the integral in eq.(\ref{aren})  perturbatively  around $z_*/z_H=1$ as follows
\begin{equation}\label{EEforddimLQzerotemp}
{\cal A}_A = \Big[\frac{2}{d-2}\left(\frac{L}{a}\right)^{d-2} + 
V\mu^{d-1}\left( \frac{d-2}{\sqrt{d(d-1)}}\right)^{d-1}
+ L^{d-2}N(b_0)\left(\frac{d-2}{\sqrt{d(d-1)}}\right)^{d-2}\mu^{d-2}\Big], 
\end{equation}
where $V=L^{d-2}l$ is the volume of the strip and $N(b_0)$ is the value of $N(\varepsilon)$ at $\varepsilon=b_0$.
The holographic entanglement negativity  for the mixed state of the adjacent subsystem of rectangular strip geometries, in the large charge limit may then be obtained utilizing our conjecture as follows
\begin{equation}\label{ENforddim4}
\mathcal{E} = \frac{3}{16 G_N^{d+1}}\Big[\frac{2}{d-2}\left(\frac{L}{a}\right)^{d-2}
+ L^{d-2}N(b_0)\left(\frac{d-2}{\sqrt{d(d-1)}}\right)^{d-2}\mu^{d-2}\Big].
\end{equation}
We observe from the above equation that in the limit of large chemical potential also the holographic entanglement negativity obtained from our conjecture is purely dependent on the area of the entangling surface shared by the
adjacent subsystems with rectangular strip geometries. As earlier for the extremal case in the $AdS_4/CFT_3$ scenario, the volume dependent contributions arising from the counting entropy of the degenerate $CFT_d$ vacuum, cancel leaving a purely area dependent expression in conformity with quantum information expectation.

\section{Summary and conclusion}\label{sec5}

To summarize, we have applied our holographic entanglement negativity conjecture for bipartite mixed state configurations of adjacent subsystems to specific examples of $CFT_d$s dual to bulk non extremal and extremal RN-$AdS_{d+1}$ black holes. Our conjecture involves a specific algebraic sum of the areas of co-dimension two bulk static minimal surfaces anchored on the appropriate subsystems in the dual $CFT_d$. In this context we have considered mixed state configurations of adjacent subsystems with rectangular strip geometries in the holographic $CFT_d$. 

In this exercise we have first studied the above examples in the $AdS_4/CFT_3$ scenario to elucidate the non trivial structure of the perturbative expansion for the holographic entanglement negativity involving various limits of the relevant parameters. For the finite temperature mixed states of adjacent subsystems in the $CFT_3$ dual to bulk non extremal RN-$AdS_{3+1}$ black hole, we observe the following behavior for the holographic entanglement negativity. In the small charge and low temperature limit the leading part of the holographic entanglement negativity includes a contribution from the zero temperature mixed state of adjacent subsystems in the $CFT_3$ which is dual to the bulk pure $AdS_4$ geometry and a correction term involving the charge and the temperature. This is because in this limit the bulk static minimal surfaces are located far away from the black hole horizon and the leading contribution to the holographic entanglement negativity arises from the near boundary pure $AdS_4$ geometry . On the other hand in the limits of large charge and low temperature and vice versa the leading part of the holographic entanglement negativity depends purely on the area of the entangling surface 
(length in $AdS_4/CFT_3$) and the volume (area in $AdS_4/CFT_3$) dependent thermal terms cancel. This is in conformity with quantum information expectation for the entanglement negativity, as the dominant contribution arises from the entanglement between the degrees of freedom at the entangling surface (line for the $AdS_4/CFT_3$ scenario) shared between the adjacent subsystems. For the case of the zero temperature mixed state of the $CFT_3$ dual to the bulk extremal RN-$AdS_{3+1}$ black hole, the leading contribution to the holographic entanglement negativity in the small charge limit consists of two parts. These involve the contribution from the zero temperature mixed state in the $CFT_3$ dual to the bulk pure $AdS_4$ geometry and a correction term involving the charge. In contrast, in the limit of large charge the leading part of the holographic entanglement negativity depends only on the area of the entangling surface ( length in $AdS_4/CFT_3$). This is due to the cancellation of the volume (area in $AdS_4/CFT_3$) dependent terms arising from the counting entropy of the degenerate $CFT_3$ vacuum.

Following the above exercise for the $AdS_4/CFT_3$ scenario to clarify the non trivial limits 
associated with the perturbation expansion, we have subsequently described the holographic entanglement negativity in the general $AdS_{d+1}/CFT_d$ case. To this end the relevant perturbative expansion of the holographic entanglement negativity requires the introduction of distinct parameters described by the energy and an effective temperature of the $CFT_d$. The leading contribution to the holographic entanglement negativity following from our conjecture exhibits identical behavior to that of the corresponding $AdS_4/CFT_3$ case for the appropriate limits of the relevant parameters. 

Our results for the applications described above, conform to the standard quantum information expectations. This may be observed from the fact that for the small chemical potential and low temperature, the contribution to the holographic entanglement negativity arises from that of the zero temperature mixed state of the $CFT_d$ dual to the bulk pure $AdS_{d+1}$ geometry
and corrections due to the chemical potential and the temperature. This conforms to the fact that in this limit the mixed state in question of the $CFT_d$ dual to the non extremal bulk RN-$AdS_{d+1}$ is dominated by the quantum correlations. For the extremal black hole on the other hand, in this limit the holographic entanglement negativity arises from that of a distinct zero temperature mixed state of the $CFT_d$ dual to the bulk pure $AdS_{d+1}$ geometry (this mixed state is obtained by tracing over the pure vacuum state of the $CFT_d$) and corrections due to the chemical potential. Once again this indicates that the mixed state above is dominated by the quantum correlations in this limit.

Furthermore our results also demonstrate the exact cancellation of the volume dependent thermal terms for the holographic entanglement negativity of the mixed state in the limit of large chemical potential and/or high temperature. Interestingly for the zero temperature mixed state of the $CFT_d$ dual to the extremal RN-$AdS_{d+1}$ black holes the cancellation involves the volume dependent counting entropy of the corresponding degenerate $CFT_d$ vacuum. Our results seemingly indicates that this is a universal feature of the holographic entanglement negativity for $CFT$s.
In both these cases the holographic entanglement negativity depends purely on the area of the entangling surface shared between the adjacent subsystems in conformity with quantum information theory expectations.

These constitute important consistency checks for the universality of our conjecture which should find interesting applications in diverse areas such as condensed matter physics and issues of quantum gravity. We should however mention here that a bulk proof for our conjecture along the lines of \cite{Lewkowycz:2013nqa} is a critical open issue which needs attention. We hope to address these fascinating issues and further applications in the near future.

\section{Acknowledgment}
Parul Jain would like to thank Prof. Mariano Cadoni for his guidance and 
the Department of Physics, Indian Institute of Technology Kanpur, India for their warm hospitality. 
Parul Jain's work is financially supported by Universit\`a di Cagliari, Italy and INFN, Sezione di Cagliari, Italy.


\begin{appendices}

\section{Non-extremal  and extremal RN-$\mathrm{AdS_4}$}\label{appenA}
 
\subsection{Non-extremal RN-$AdS_4$ (Small charge - high temperature)}\label{appen1}

The constants  $k_1$, $k_2$, $k_3$, $k_4$ and $k_5$  in the eq. (\ref{shight}) are given as   follows
\begin{eqnarray}\label{constants}
k_1 &=& \sum_{n=1}^\infty\bigg(\frac{1}{3n-1} \frac{\Gamma(n+\frac{1}{2})}
{\Gamma(n+1)}\frac{\Gamma(\frac{3n+3}{4})}{\Gamma(\frac{3n+5}{4})}-\frac{2}{3\sqrt{3}n^2}\bigg) + \frac{\pi^2}{9\sqrt{3}}
+\frac{\sqrt{\pi} \Gamma(-\frac{1}{4})}{\Gamma(\frac{1}{4})},\label{k1}\\
k_2 &=& \frac{3\pi}{8}-\frac{3 \Gamma(\frac{3}{2})\Gamma(\frac{7}{4})}{\Gamma(\frac{9}{4})}+3\sum_{n=1}^\infty\bigg(\frac{1}{3n+2} \frac{\Gamma(n+\frac{3}{2})}{\Gamma(n+1)}\frac{\Gamma(\frac{3n+6}{4})}{\Gamma(\frac{3n+8}{4})}-\frac{1}{3\sqrt{3}n}\bigg)\nonumber\\
&-& 3\sum_{n=1}^\infty\bigg(\frac{2}{3n+3} \frac{\Gamma(n+\frac{3}{2})}{\Gamma(n+1)}\frac{\Gamma(\frac{3n+7}{4})}{\Gamma(\frac{3n+9}{4})}-\frac{2}{3\sqrt{3}n}\bigg)\label{k2},\\
k_3&=&\frac{-2}{\sqrt{3}} + \frac{\pi^2}{9\sqrt{3}},\label{k3}\\
k_4 &=& \frac{2}{\sqrt{3}}-\frac{2}{\sqrt{3}}\log[3]+\frac{3\sqrt{\pi}\Gamma(\frac{3}{2})\Gamma(\frac{7}{4})}{\Gamma(\frac{9}{4})},\label{k4}\\ 
k_5 &=& \frac{-2}{\sqrt{3}}\label{k5}.
\end{eqnarray}
The constants $c_1$ and $c_2$ appearing  in  the eq. (\ref{epsilon})  are given as follows
\begin{eqnarray}
 c_1 &=& \frac{\sqrt{\pi}}{2}\frac{\Gamma(\frac{3}{4})}{\Gamma(\frac{5}{4})}+ \sum_{n=1}^\infty\bigg( \frac{\Gamma(n+\frac{1}{2})}{2\Gamma(n+1)}\frac{\Gamma(\frac{3n+3}{4})}{\Gamma(\frac{3n+5}{4})}-\frac{1}{\sqrt{3}n}\bigg),\label{c1}\\
  c_2 &=& \frac{1}{\sqrt{3}}-\frac{3}{2}\sum_{n=0}^\infty \bigg( \frac{\Gamma(n+\frac{3}{2})}{\Gamma(n+1)}\frac{\Gamma(\frac{3n+6}{4})}{\Gamma(\frac{3n+8}{4})}-\frac{2}{\sqrt{3}}\bigg) +\frac{3}{2}\sum_{n=0}^\infty\bigg( \frac{\Gamma(n+\frac{3}{2})}{\Gamma(n+1)}\frac{\Gamma(\frac{3n+7}{4})}{\Gamma(\frac{3n+9}{4})}-\frac{2}{\sqrt{3}}\bigg).\label{c2}
\end{eqnarray}

 \subsection{Non-extremal RN-$\mathrm{AdS_4}$ (Large charge - high temperature)}\label{appen2}

 The constants $K'_1$ and $K'_2$  in the eq. (\ref{EEforLQLTemp}) are given as follows
\begin{eqnarray}
K_1'&=&-\frac{2\sqrt{\pi}\Gamma(\frac{3}{4})}{\Gamma(\frac{1}{4} )}+\frac{\log[4]-10}{8} +\frac{1}{2}\sum_{n=2}^\infty\bigg(\frac{1}{n-1}\frac{\Gamma(n+\frac{1}{2})}{\Gamma(n+1)}\frac{\Gamma(\frac{n+3}{4})}{\Gamma(\frac{n+5}{4})}-\frac{2}{n^2}\bigg)+\frac{\pi^2}{6},\label{K'1}\\
K_2'&=& \frac{\pi^2 }{6}- \frac{3}{2}\label{K'2}.
\end{eqnarray}

\subsection{Extremal RN-$\mathrm{AdS_4}$ (Large charge )}\label{appen3}

The constants $K_1$, $K_2$ and $K_3$ in the eq. (\ref{EEforLQExtr}) are given as follows
\begin{eqnarray}
K_1 &=& \frac{2}{\sqrt{6}}\bigg[-2\frac{\sqrt{\pi } \Gamma (\frac{3}{4})}{ \Gamma (\frac{1}{4})}+\frac{\log[4]}{4}-\frac{1+2\sqrt{\pi}}{2}+ \sqrt{\pi } \zeta \left(\frac{3}{2}\right)\nonumber\\
&+&\frac{\sqrt{\pi }}{2}\sum^\infty_{n=2}
(\frac{1}{n-1}\frac{\Gamma (\frac{n+3}{4})}{ \Gamma (\frac{n+5}{4})}-\frac{2}{n\sqrt{n}})\bigg],\label{K1} \\
K_2 &=& -\frac{2\pi}{\sqrt{6}},\label{K2} \\
K_3&=&\frac{2}{\sqrt{6}}\bigg[1-\sqrt{\pi}+
\sqrt{\pi }  \zeta \left(\frac{3}{2}\right)\bigg].\label{K3}
\end{eqnarray}


\section{Non-extremal  and extremal RN-$\mathrm{AdS_{d+1}}$ }\label{B}

\subsection{Non-extremal RN-$\mathrm{AdS_{d+1}}$ (Small chemical potential - low temperature)}\label{B1}

The constants $\mathcal{S}_0$ and $\mathcal{S}_1$ appearing in the eq. (\ref{EEforddimLQlowtemp}) are given as follows
\begin{align}
\mathcal{S}_0=& \frac{2^{d-2} \pi ^{\frac{d-1}{2}} \Gamma 
\left(-\frac{d-2}{2 (d-1)}\right) }{(d-1) \Gamma \left(\frac{1}{2 (d-1)}\right)} \left(\frac{\Gamma
\left(\frac{d}{2 (d-1)}\right)}{\Gamma \left(\frac{1}{2 (d-1)}\right)}\right)^{d-2}, \label{s0} \\
\mathcal{S}_1=& \frac{\Gamma \left(\frac{1}{2 (d-1)}\right)^{d+1}2^{-d-1} 
\pi ^{-\frac{d}{2}}}{\Gamma \left(\frac{d}{2(d-1)}\right)^d\Gamma
\left(\frac{1}{2}+\frac{1}{d-1}\right)} \left(\frac{\Gamma \left(\frac{1}{d-1}\right) }{\Gamma \left(-\frac{d-2}{2 (d-1)}\right)}+\frac{2^{\frac{1}{d-1}} (d-2)
\Gamma \left(1+\frac{1}{2 (d-1)}\right) }{\sqrt{\pi } (d+1)}\right).\label{s1}
\end{align}

\subsection{Non-extremal RN-$\mathrm{AdS_{d+1}}$ (Small chemical potential - high temperature)}\label{B2}

The function $\gamma_d\Big(\frac{\mu}{T}\Big)$ appearing in the eq. (\ref{EEforddimSQhightemp}) is given as follows
\begin{equation}
\gamma_d\left(\frac{\mu}{T}\right)=N(1)+\frac{d^2(d-2)}{16\pi^2(d-1)}
\left(\frac{\mu}{T}\right)^2\int_{0}^{1}dx\left(\frac{x\sqrt{1-x^{2(d-1)}}}{\sqrt{1-x^d}}\right)
\left(\frac{1-x^{d-2}}{1-x^d}\right)+\mathcal{O}\left(\frac{\mu}{T}\right)^4,\label{gammad}
\end{equation}
where the numerical constant $N(\varepsilon)$ is given as
\begin{equation}
N(\varepsilon)=2\left[\frac{\sqrt{\pi } \Gamma \left(-\frac{d-2}{2 (d-1)}\right)}{2 (d-1) \Gamma \left(\frac{1}{2
(d-1)}\right)}\right]+2\int_{0}^{1}dx
\left(\frac{\sqrt{1-x^{2(d-1)}}}{x^{d-1}\sqrt{f(z_H x)}}-\frac{1}{x^{d-1}\sqrt{1-x^{2(d-1)}}}\right).\label{Nepsilon}
\end{equation}

\subsection{Non-extremal RN-$\mathrm{AdS_{d+1}}$ (Large chemical potential - high temperature)}\label{B3}

The numerical constants $ N_{0},N_{1}$ in the eq. (\ref{EEforddimLQhightemp}) are given as follows
\begin{align}
N_0&=2\left[\frac{\sqrt{\pi } \Gamma \left(-\frac{d-2}{2 (d-1)}\right)}{2 (d-1) \Gamma \left(\frac{1}{2
(d-1)}\right)}\right]+2\int_{0}^{1}dx\left(\frac{\sqrt{1-x^{2(d-1)}}}{x^{d-1}
\sqrt{1-b_0 x^d+b_1 x^{2(d-1)}}}-\frac{1}{x^{d-1}\sqrt{1-x^{2(d-1)}}}\right),\label{N0} \\
N_1&=\int_{0}^{1}dx\left(\frac{x\sqrt{1-x^{2(d-1)}}}{\sqrt{1-b_0 x^d+b_1 x^{2(d-1)}}}\right)
\left(\frac{1-x^{d-2}}{1-b_0 x^d+b_1 x^{2(d-1)}}\right).\label{N1}
\end{align}

\end{appendices}

\bibliographystyle{utphys}
\bibliography{ENforRN}

\end{document}